\documentclass[a4paper,11pt]{article}
\pdfoutput=1 
\usepackage{amsmath,times,amssymb,latexsym,multicol,graphics}
\usepackage{ascmac,fancybox}
\usepackage{bbm}
\usepackage{braket}
\newcommand{\be}{\begin{equation}}
\newcommand{\ee}{\end{equation}}

\newcommand{\D}{\Delta}

\newcommand{\g}{\gamma}

\newcommand{\la}{\lambda}

\newcommand{\p}{\partial}

\usepackage{fullpage}
\begin{document} 
\begin{titlepage}
\begin{flushright}
{\small OU-HET 965}
 \\
\end{flushright}

\begin{center}

\vspace{1cm}

\hspace{3mm}{\LARGE \bf Towards Entanglement of Purification\\ 
for Conformal Field Theories
} \\[3pt] 
\vspace{1cm}

\renewcommand\thefootnote{\mbox{$\fnsymbol{footnote}$}}
Hayato {Hirai}${}^1$, Kotaro {Tamaoka}${}^2$, Tsuyoshi {Yokoya}${}^3$

\vspace{5mm}

{\small \sl Department of Physics, Osaka University} \\ 
{\small \sl Toyonaka, Osaka 560-0043, JAPAN}

\vspace{5mm}
${}^1${\small{\,hirai@het.phys.sci.osaka-u.ac.jp}
}\\
${}^2${\small{\,k-tamaoka@het.phys.sci.osaka-u.ac.jp}
}\\
${}^3${\small{\,yokoya@het.phys.sci.osaka-u.ac.jp}
}
\end{center}

\vspace{5mm}

\noindent
\abstract
We argue that the entanglement of purification for two dimensional holographic CFT can be obtained from conformal blocks with internal twist operators. First, we explain our formula from the view point of tensor network model of holography. Then, we apply it to bipartite mixed states dual to subregion of AdS${}_3$ and the static BTZ blackhole geometries. The formula in CFT agrees with the entanglement wedge cross section in the bulk, which has been recently conjectured to be equivalent to the entanglement of purification. 

\end{titlepage}
\setcounter{footnote}{0}
\renewcommand\thefootnote{\mbox{\arabic{footnote}}}
\tableofcontents
\flushbottom

\section{Introduction}
\ \ \ \ In the AdS/CFT duality\cite{Maldacena:1997re}, bulk geometries have profound connection to the quantum correlations in the conformal field theories (CFT). 
To deepen our understanding of this interesting connection, it would be important to reveal which kind of correlation in CFT corresponds to a given geometrical object in the bulk. 
The most well-known example is the equivalence between the area of minimal surface in AdS and the entanglement entropy (EE) in CFT\cite{Ryu:2006bv}. 
For the mixed states, however, EE is not very nice measure for the quantum entanglement because it also picks up the thermal entropy. There has been several attempts to find measure that captures just quantum correlation for mixed states.

One of such measure is entanglement of purification (EoP)\cite{Terhal:2002} which is a measure for correlation in a given bipartite mixed state. In general, we can purify a mixed state $\rho_{AB}$ on a Hilbert space $\mathcal{H}_{A}\otimes\mathcal{H}_{B}$ to a pure state $\ket{\psi}$ on an enlarged Hilbert space $\mathcal{H}_{A}\otimes\mathcal{H}_{B}\otimes\mathcal{H}_{A'}\otimes\mathcal{H}_{B'}$. There are infinitely many ways of purification $\ket{\psi}$ such that $\rho_{AB}=\textrm{Tr}_{A^\prime B^\prime}\ket{\psi}\bra{\psi}$. 
For a given bipartite mixed state $\rho_{AB}$, EoP $E_P(A:B)$ is defined by minimum EE for all possible purifications;
\be
E_P(A:B)=\underset{\rho_{AB}=\textrm{Tr}_{A^\prime B^\prime}\ket{\psi}\bra{\psi}}{\textrm{min}}S(\rho_{AA^\prime}), 
\ee
where $\rho_{AA^\prime}=\textrm{Tr}_{BB^\prime}\ket{\psi}\bra{\psi}$ and $S(\rho_{AA^\prime})$ is EE associated with $\rho_{AA^\prime}$. 
Note that for the pure states EoP is equivalent to the EE since we do not need purification. It is hard to calculate EoP in practice because one needs to find an optimized solution, which minimizes the $S(\rho_{AA^\prime})$ displayed above, from all possible purifications. In fact, there are a few example calculating the EoP in many body system\cite{Nguyen:2017yqw,Bhattacharyya:2018sbw} and no example directly from the definition in quantum fields theory.

In this paper, we propose a formula of EoP for two dimensional holographic CFT\cite{Heemskerk:2009pn,ElShowk:2011ag} by using the replica trick\cite{Callan:1994py}, as well as EE. 
As always, we can apply this trick for EE with respect to a purified state that minimizes EE; we will call it an optimized solution $\ket{\Psi_{opt.}}$, 
\be
E_P(A:B)=S(\rho^{(opt.)}_{AA^\prime})=-\textrm{Tr}\rho^{(opt.)}_{AA^\prime}\log\rho^{(opt.)}_{AA^\prime}=\left.-\dfrac{\p}{\p n}\textrm{Tr}\left(\rho^{(opt.)}_{AA^\prime}\right)^n\right|_{n\rightarrow1},
\ee
where $\rho^{(opt.)}_{AA^\prime}=\textrm{Tr}_{BB^\prime}\ket{\Psi_{opt.}}\bra{\Psi_{opt.}}$. Once entangling surfaces (i.e. boundary points between $A'$ and $B'$) are specified, one may further write it in terms of correlation function of (external) twist operators\cite{Calabrese:2004eu}. We argue that $\textrm{Tr}(\rho^{(opt.)}_{AA^\prime})^n$ for holographic CFT can be well approximated as the Virasoro conformal blocks, including twist operators as intermediate state, 
\be
E_P(A:B)=-\left.\dfrac{\p}{\p n}\mathcal{F}_{\D_n}\right|_{n\rightarrow1}.\label{eq:eopcb}
\ee
More information about the blocks $\mathcal{F}_{\D_n}$ will be explained in the following section. If one considers a mixed state, associated with the subregion of the vacuum state, the Virasoro conformal blocks further reduce to the global conformal blocks. 
In section \ref{sec:HaPPY}, we give an interpretation of \eqref{eq:eopcb} from the view point of a tensor network model of Holography\cite{Pastawski:2015qua}. 
Based on the insight from this model, in section \ref{subsec:ads3}, we apply our formula \eqref{eq:eopcb} for the aforementioned mixed state. Moreover, we also consider the EoP for the thermal state which is dual to the BTZ blackhole\cite{Banados:1992wn} in section \ref{subsec:btz}. 

In particular, our computation in section \ref{sec:ew} agrees with an interesting conjecture which recently proposed by \cite{Takayanagi:2017knl,Nguyen:2017yqw} (see also \cite{Bao:2017nhh}). They considered the minimal cross section of entanglement wedge $\sigma_{min.}$ and defined a quantity ``entanglement wedge cross section'' by
\be
E_W=\dfrac{\sigma_{min.}}{4G_N}, 
\ee
where $G_N$ is the Newton's constant. Their claim is that $E_P=E_W$ for CFT with the bulk dual at the leading order of large-$c$ (large-$N$) expansion. 
Our argument in section \ref{sec:ew} gives a derivation of $E_W$ within the framework of CFT. 
We discuss the implication of our formula and future direction in section \ref{sec:discussion}.

\section{Some implications to EoP in AdS/CFT from holographic code model}\label{sec:HaPPY}

\ \ \ \ This section describes an heuristic justification for \eqref{eq:eopcb} by using the holographic code model\cite{Pastawski:2015qua}. 
In section \ref{subsec:eopewc}, we evaluate the EoP in the model and see the $E_P=E_W$ conjecture is actually satisfied. Then in section \ref{subsec:eopewc}, based on the exact result of optimized purification, we argue the EoP could be calculated in terms of the bulk two point functions on geodesics. We will identify the two point function with the conformal blocks in the next section.


\subsection{Interpretation of $E_P$ = $E_W$ conjecture in holographic code model}\label{subsec:eopewc}

\ \ \ \ The holographic code model is a toy model of the AdS/CFT duality constructed by the tensor network. It captures the important relations between the bulk geometry and entanglement structures of the boundary theory; for example, Ryu-Takayanagi formula and quantum error correction feature about the boundary dual of bulk local operators in the low energy states (HKLL bulk reconstruction\cite{Hamilton:2006az,Almheiri:2014lwa}). Thus, we expect that considering the interpretation of the $E_P=E_W$ conjecture in holographic code model would help to promote our understanding of the conjecture in the AdS/CFT duality.   

In holographic code model, the duality map from bulk states to boundary states is an isometry map constructed by putting the so-called perfect tensors on the uniformly tiled two dimensional hyperbolic space. This isometry map is known as ``holographic code". The features of perfect tensor ensures that this code is just a quantum error correcting code which embeds the bulk Hilbert space into boundary one. The non-uniqueness of the reconstruction of the bulk local operators in the boundary theory can also be understood as the well known property of the quantum error correcting code against erasure errors. Therefore, we think of this model as a toy model of the low energy sector of the AdS/CFT with classical geometries. Then, we assume that the bulk state is the vacuum state and the quantum bulk degrees of freedoms(d.o.f) of the tensor correspond to the d.o.f of the quantum fields in the semiclassical theory. Although Hamiltonian is not specified in this model, we consider the bulk vacuum state as a product state so that the Ryu-Takayanagi formula holds without quantum corrections\footnote{In the holographic code model, the HKLL-like property holds independently from the bulk state. On the other hand, if the bulk state is entangled, the Ryu-Takayanagi formula get quantum corrections, which is the EE of bulk state.}.

\begin{figure}[tbp]
\begin{center}
\resizebox{170mm}{!}{\includegraphics{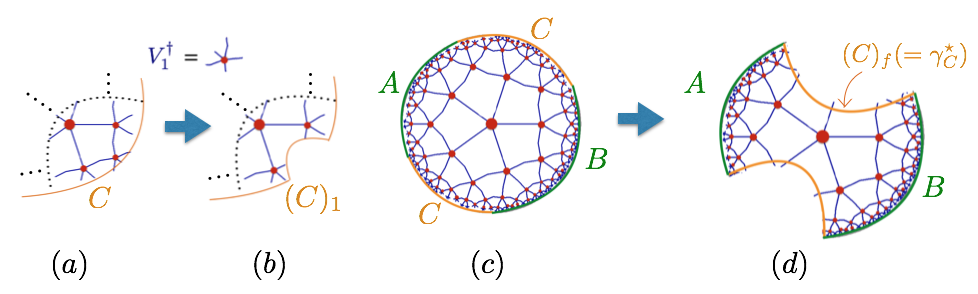}}
\caption{Figure(a) represents a part around the boundary within the tensor network(TN) of $|0\rangle$. Figure(b) is the TN representation of $|\Psi\rangle_{AB(C)_{1}}$ which is the state cut a perfect tensor from $|0\rangle$. Figure(c) represents the whole TN of $|0\rangle$ and Figure(d) is the optimally purified state $|\Psi_{opt}\rangle$ constructed by the iteratively cutting the perfect tensors. }\label{fig:opt}
\end{center}
\end{figure} 

At first, we discuss the EoP in this model. Let us consider a bipartite mixed state $\rho_{AB}$ in the boundary theory which is obtained by tracing out the d.o.f on the boundary region $C(=\overline{AB})$ (see Figure \ref{fig:opt}.($c$)). To get the EoP of $\rho_{AB}$, we purify $\rho_{AB}$ at first.  There are many possible ways of the purification, but the one of them which minimizes the EE of  $\rho_{AA'}$ must be chosen to get the EoP. It seems very difficult to choose such an optimized purification in practice, but in the holographic code model, it can be done very easily. The calculation of the EoP of $\rho_{AB}$ in this model is implemented by the following steps. 
\begin{enumerate}
\item Choose the original boundary pure state $|\Psi\rangle_{ABC}$, which is supposed to be the vacuum state in AdS/CFT, as the initial purification to $\rho_{AB}$. This state can be represented as the tensor network (c) of Figure \ref{fig:opt}. In this step, the Hilbert space associating with the d.o.f on the boundary subregion $C$ which is complement to $A\cup B$ is added to Hilbert space $\mathcal{H}_{AB}$ as a purification. 

\item Act the hermitian conjugate of  isometry matrix $V^{\dagger}_{1}$ which cuts the perfect tensor sitting on $C$:
\be \label{psi1:eq}
V^{\dagger}_{1}|\Psi\rangle_{ABC}=|\Psi_{1}\rangle_{AB(C)_{1}},
\ee
where the $AB(C)_{1}$ is a new boundary obtained by removing the portion cut by the $V^{\dagger}_{1}$ from bulk region which has $ABC$ as its boundary(see (a) and (b) of Figure \ref{fig:opt}). Note that in this process, $|\Psi_{1}\rangle_{AB(C)_{1}}$ become the new purification of $\rho_{AB}$ as follows:
\be
{\mathop{\mathrm{Tr}}_{AB(C)_{1}}}\Big{[} |\Psi_{1}\rangle\langle\Psi_{1}|\Big{]}={\mathop{\mathrm{Tr}}_{AB(C)_{1}}} \Big{[}V^{\dagger}V|\Psi_{1}\rangle\langle\Psi_{1}|\Big{]}
={\mathop{\mathrm{Tr}}_{ABC}} \Big{[}|\Psi\rangle\langle\Psi|\Big{]}
=\rho_{AB},\label{trace:eq}
\ee
where we used $V^{\dagger}V=I$, which is the property of the isometry matrix, in the first equality and cyclic property of trace and (\ref{psi1:eq}) in the second equality
\footnote{Precisely, more effort is needed to justify the second equality in (\ref{trace:eq}) since the size of the Hilbert space $\mathcal{H}_{ABC}$ and $\mathcal{H}_{AB(C)_{1}}$ are not the same. To make them same, we add  auxiliary d.o.fs $|0\rangle_{C'_{1}} \in \mathcal{H}_{C'_{1}}$ to $|\Psi_{1}\rangle$ such that  $\mathcal{H}_{ABC}=\mathcal{H}_{AB(C)_{1}}\otimes\mathcal{H}_{C'_{1}}$. Then, both $|\Psi_{1}\rangle|0\rangle_{C'_{1}}$ and $|\Psi\rangle$ are an state in $\mathcal{H}_{ABC}$. Then, there exist the unitary operator $U:\mathcal{H}_{ABC}\rightarrow \mathcal{H}_{ABC}$ for given $V$ such that $V=U|0\rangle_{C'_{1}}$ where $V$ is an isometry map from $\mathcal{H}_{AB(C)_{1}}$ to  $\mathcal{H}_{ABC}$. Therefore,
\be
{\mathop{\mathrm{Tr}}_{AB(C)_{1}}}\Big{[} |\Psi_{1}\rangle\langle\Psi_{1}|\Big{]}
={\mathop{\mathrm{Tr}}_{ABC}}\Big{[} |\Psi_{1}\rangle|0\rangle\langle0|\langle\Psi_{1}|\Big{]}
={\mathop{\mathrm{Tr}}_{ABC}} \Big{[}U^{\dagger}U|\Psi_{1}\rangle|0\rangle\langle0|\langle\Psi_{1}|\Big{]}
={\mathop{\mathrm{Tr}}_{ABC}} \Big{[}|\Psi\rangle\langle\Psi|\Big{]}
=\rho_{AB}.
\ee }.
Then, we can iterate this procedure by acting the hermitian conjugate of  isometry matrix $V^{\dagger}_{i}$ to $|\Psi_{i}\rangle_{AB(C)_{i}}$ which maps from $\mathcal{H}_{AB(C)_{i}}$ to $\mathcal{H}_{AB(C)_{i+1}}$:
\be
|\Psi_{i}\rangle_{AB(C)_{i}}\  \rightarrow\  |\Psi_{i+1}\rangle_{AB(C)_{i+1}}=V^{\dagger}_{i}|\Psi_{i}\rangle_{AB(C)_{i}},
\ee
where $\mathcal{H}_{AB(C)_{i+1}}$ is defined from $\mathcal{H}_{AB(C)_{i}}$ in the same way as $\mathcal{H}_{AB(C)_{1}}$. Since the hermitian conjugate of the isometry maps reduce the size of Hilbert space, this procedure reduces the size of the purified Hilbert space. Thus, the entanglement of purification monotonically decreases. The boundary legs of the state $|\Psi_{i}\rangle$ enter deep into the bulk as the size of the purified Hilbert space reduces. This iterative procedure will terminate when the $V^{\dagger}_{f}$ acting on the  $\mathcal{H}_{AB(C)_{f}}$ no longer exists. 

\item After the procedure (ii) ends, divide the boundary subregion $(C)_{f}$ into two parts, $A'$ and $B'$ so that $A'$ and $B'$ adjacent to the $A$ and $B$ respectively. Then, we can obtain the density matrix $\rho_{AA'}$ by tracing out $\mathcal{H}_{BB'}$ from $|\Psi_{f}\rangle_{ABA'B'}$ and calculate the EE of $\rho_{AA'}$.

\item Iterate the procedure (iii) with different division of $(C)_{f}$ in order to find the optimized division, say $A'_{min}B'_{min}$, which minimizes the EE of $\rho_{AA'}$. Then, that minimum EE  gives the EoP of $\rho_{AB}$.

\end{enumerate}    

\begin{figure}[tbp]
\begin{center}
\resizebox{70mm}{!}{\includegraphics{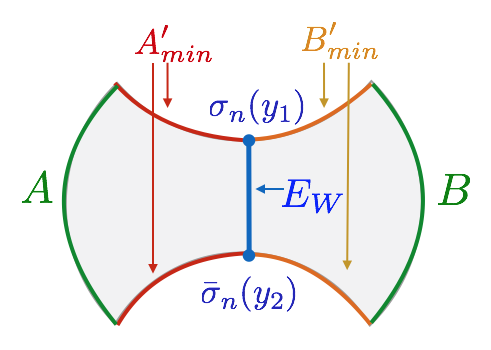}}
\caption{The boundary space along  $\gamma^{\star}_{C}$ are divided into $A'_{min}$(red line) and $B'_{min}$(orange line).  The $y_{1}$ and $y_{2}$ are the two  boundary points between  $A'_{min}$ and $B'_{min}$. This division is determined so that the length of the geodesic(blue line) between $y_{1}$ and $y_{2}$ become shortest. The $E_W$ is given by that length times $1/4G_{N}$. }
\label{fig:Ew}
\end{center}
\end{figure}

It is very important that the each steps of iterative procedure (ii) exactly corresponds to steps of the ``greedy algorithm'' defined in \cite{Pastawski:2015qua} which is the algorithm to obtain the geodesics on the discretized hyperbolic space. The geodesics obtained by the algorithm are called greedy geodesics. The $A'_{min}B'_{min}$  end up with $\gamma^{\star}_{C}$ which is the greedy geodesic with $\partial C$ as its boundary, namely $\partial{\gamma^{\star}_{C}}=\partial C$.
Then, in the step (iii), given the division of $(C)_{f}$ into $A'$ and $B'$, the EE of $\rho_{AA'}$ is given by the length of the $\gamma^{\star}_{BB'}$. The optimized division is the one with minimum length of $\gamma^{\star}_{BB'}$. Moreover, this optimized  $\gamma^{\star}_{BB'}$ is exactly the entanglement wedge cross section of  $\rho_{AB}$ (Figure \ref{fig:Ew}).
Eventually, $E_P=E_W$ conjecture is correct in the holographic code model.

\subsection{EoP from a two point function in the bulk}\label{subsec:2pt}

As a result of the previous subsection, optimally purified state $|\Psi_{opt}\rangle\equiv |\Psi_{f}\rangle_{ABA'_{min}B'_{min}}$, which is the pure state in the Hilbert space associated with $ABA'_{min}B'_{min}$, can be written as $|\Psi_{opt}\rangle=V^{\dagger}|0\rangle$ where $V=V_{1}V_{2}\cdots V_{f}$. Although this is the result in holographic code model, let's imagine the $|\Psi_{opt}\rangle$ in  AdS/CFT. Namely, the bulk geometry is cut along the geodesic $\gamma_{C}$ as in Figure \ref{fig:Ew} and boundary d.o.f live on $ABA'_{min}B_{min}'$. Then $E_{P} (A:B)$ would be calculated by using the replica method as 
\be\label{Ep:eq}
E_P(A:B)=-\dfrac{\p}{\p n} \langle\Psi_{opt}|\sigma_{n}(y_1)\bar{\sigma}_{n}(y_{2})|\Psi_{opt}\rangle \Big{|}_{n\rightarrow1},
\ee
where $\sigma_{n}(y)$ and $\bar{\sigma}_{n}(y)$ are the twist operators acting on the $n$-sheeted boundary space along $\gamma_{C}$, not original boundary on which CFT lives. The $y_{1}$ and $y_{2}$ are the boundary points between $A'_{min}$ and $B'_{min}$(Figure \ref{fig:Ew}). However, we cannot compute (\ref{Ep:eq}) directly since we don't know what exactly is the $|\Psi_{opt}\rangle$  in AdS/CFT. Thus, we need to put (\ref{Ep:eq}) into the original CFT language. This can be done by using the $|\Psi_{opt}\rangle=V^{\dagger}|0\rangle$:
\begin{align}
E_P(A:B)&=-\dfrac{\p}{\p n} \langle0|V\sigma_{n}(y_1)V^{\dagger}V\bar{\sigma}_{n}(y_{2})V^{\dagger}|0\rangle \Big{|}_{n\rightarrow1}\nonumber\\
          &=-\dfrac{\p}{\p n} \langle0|(K\sigma_{n})(y_1)(K\bar{\sigma}_{n})(y_{2})|0\rangle \Big{|}_{n\rightarrow1},\label{Epincft:eq}
\end{align}
where we defined $(K\mathcal{O})(y)\equiv V\mathcal{O}(y)V^{\dagger}$. Although we do not know the concrete expression of $V$ in AdS/CFT,   the $(K\mathcal{O})(y)$ is the boundary operator which is dual to the bulk local operator $\mathcal{O}(y)$ in the holographic code model\footnote{Strictly speaking, in the holographic code model, $(K\mathcal{O})(y)$ is not the boundary operator dual to bulk operator $\sigma^{(bulk)}_{n}(y)$ but the boundary operator dual to the $\sigma_{n}(y)$, acting on the boundary cut along  $\gamma^{\star}_{C}$. The $\sigma_{n}(y)$ and $\sigma^{(bulk)}_{n}(y)$ are acting on a bulk leg and a boundary leg of the same perfect tensor on $\gamma^{\star}_{C}$ , respectively. We need $\sigma_{n}(y)$ acting on at least three legs of the perfect tensor to reconstruct the bulk local(1-body) operator $\sigma^{(bulk)}_{n}(y)$, which should be understood as discretized version of the bulk reconstruction. However, we do not distinguish between them since these operators are identified up to scaling factor in AdS/CFT.}. 
Therefore, it would be natural to think of the $(K\mathcal{O})(y)$ in  AdS/CFT as  non-local operator on the boundary along $C$  which is constructed by the HKLL bulk reconstruction of the bulk local operator $\mathcal{O}^{(bulk)}(y)$. Thus, in the bulk language  (\ref{Epincft:eq}) can be written as
\be\label{Eopinbulk:eq}
E_P(A:B)=-\dfrac{\p}{\p n}\langle0_{AdS}|\sigma^{(bulk)}_{n}(y_1)\bar{\sigma}^{(bulk)}_{n}(y_{2})|0_{AdS}\rangle \Big{|}_{n\rightarrow1},
\ee
where $\sigma_{n}^{(bulk)}$ and $\bar{\sigma}_{n}^{(bulk)}$  are the bulk dual of the twist operators $\sigma_{n}$ and $\bar{\sigma}_{n}$ respectively with their mass $m^{2}=\Delta_{n}(\Delta_{n}-2)$. Here, $\D_n\,(\bar{\D}_n)$ is scaling dimension of the $\sigma_n\,(\bar{\sigma}_n)$. In the next section, we will see that the (\ref{Eopinbulk:eq}) agrees with the $E_W$ of $AB$ in the large-$c$ limit and this can be expressed in terms of the global conformal block in the CFT.

\section{Entanglement wedge cross section from Holographic CFT}\label{sec:ew}
\ \ \ \ In this section, we argue that the right hand side of \eqref{Eopinbulk:eq} can be obtained from conformal blocks (CBs) including twist operators as intermediate states. Having the previous argument in section \ref{sec:HaPPY}, we can regard it as EoP for holographic CFT. CBs are the basis of correlation function in CFT. These are determined purely from the conformal symmetry and irreducible representation thereof. We have an interesting integral representation of the CB, dubbed the geodesic Witten diagram (GWD)\cite{Hijano:2015zsa,Hijano:2015qja}. Remarkably, one can apply this representation for arbitrary CFT even with no bulk dual. Taking the large-$c$ limit, we can read off the entanglement wedge cross section from GWD as discussed below.

\subsection{Cross section of AdS${}_3$ from global conformal block}\label{subsec:ads3}
Firstly, we extract the right hand side of \eqref{Eopinbulk:eq} in AdS${}_3$ from CB with internal twist operators. 
Let us consider CB associated with the following 4pt function,
\be
\braket{0|\mathcal{O}_{1L}(\phi_1)\mathcal{O}_{2L}(\phi_2)\mathcal{O}_{3L}(\phi_3)\mathcal{O}_{4L}(\phi_4)|0}. \label{eq:4pt}
\ee
Hereafter we will assume that the scaling dimension of $\mathcal{O}_{iL}$s$\,(i=1,2,3,4)$ are the same. We will consider the specific channel that $\mathcal{O}_{1L}\mathcal{O}_{4L}$ fuses into $\sigma_n$ and $ \mathcal{O}_{2L}\mathcal{O}_{3L}$ does into $\bar{\sigma}_n$. Since we apply the replica trick, we are not considering the original CFT $\mathcal{C}$ but its cyclic orbifold $\mathcal{C}^n/\mathbb{Z}_n$. In the large-c holographic CFT, the contribution of the channel mentioned above will be dominant for original four point function \eqref{eq:4pt} since the conformal dimension of twist operators is the lowest one in the sector with twist number $\pm 1$ and its spectrum is sparse\footnote{In unitary CFT, the twist operators have the lowest scaling dimension in the sectors with twist number $\pm 1$ ; hence, these are the ``vacuum" states on the sectors via the operator/state map. Moreover, the other primary states in these sectors are systematically built from the primary operators in the original CFT $\mathcal{C}$ acting on these ``vacua"\cite{Borisov:1997nc}.}. The more detailed properties of $O_{iL}$ will be discussed in section \ref{sec:discussion}. 
For a while, we discuss the global CB $G_{\D_n}$ associated with the twist operators, but it turns out to be the same as Virasoro CB $\mathcal{F}_{\D_n}$ at the semi-classical limit below.
We shall consider CFT${}_2$ on the cylinder for simplicity. Then, in the bulk side, it is dual to global AdS${}_3$,
\be
ds^2=\dfrac{1}{\cos^2\rho}\left(d\rho^2-dt^2+\sin^2\rho d\phi^2\right).\label{eq:gads3}
\ee
We have a boundary cylinder at $\rho=\frac{\pi}{2}$ on which CFT${}_2$ lives. 
On the fixed time slice $t=0$, GWD is given by\footnote{Note that we are not discussing the conformal partial waves but rather the conformal blocks.}
\be
G_{\D_n}(u,v)=\int_{\g_{14}} d\la\int_{\g_{23}} d\la^\prime G^{\D_n}_{bb}(y(\la),y^\prime(\la^\prime)),\label{eq:gwdt}
\ee
where $G^{\D_n}_{bb}(y,y^\prime)$ is the scalar bulk-bulk propagator in AdS${}_3$ with mass $m^2=\D_n(\D_n-2)$. The explicit form of the propagator is given in appendix \ref{app:A}. In general, GWD includes bulk-boundary propagator. However, under our assumption about scaling dimension of the external operators, it reduces to \eqref{eq:gwdt}. This is because CB depends only on the difference of external scaling dimensions for each OPE. 
Each end points $y,y^\prime$ in \eqref{eq:gwdt} is sitting on each geodesics $\g_{14}$ and $\g_{23}$. Here we denote $\g_{ij}$ as the geodesic anchored on the boundary points $\phi_i$ and $\phi_j$. 

Let us take the semiclassical limit, that is, large-$c$ limit leaving $\D_n/c$ fixed. We will apply the following argument to the twist operators $\sigma_n, \bar{\sigma}_{n}$ with scaling dimension $\D_n=\frac{c}{12}(n-\frac{1}{n})$. In this limit, we can use the saddle point approximation for the integrand. At the leading order of the approximation, \eqref{eq:gwdt} reduces to
\be
G_{\D_n}(u,v)=e^{-\D_n\sigma_{min}(u,v)}, \label{eq:dom}
\ee
where $\sigma_{min}$ is minimum length between two geodesics. For the explicit form of  $\sigma_{min}$, see appendix \ref{app:A}. $\sigma_{min}$ is determined purely from the cross ratios, as CB does; hence, it is conformally invariant. 
Notice that the dominant contribution in the approximation \eqref{eq:dom} obviously comes from the bulk-bulk propagator stretched over two geodesics such that it minimizes the distance between two end points. Moreover, under the limit $n\rightarrow1$, we can regard $\sigma_n, \bar{\sigma}_n$ as the light operators. Then, the difference between the global CB $G_{\D_n}$ and the Virasoro one $\mathcal{F}_{\D_n}$ becomes negligible\cite{Fitzpatrick:2014vua}. Therefore, we can identify the Virasoro CB $\mathcal{F}_{\D_n}$ with the two point function in \eqref{Eopinbulk:eq},
\be
\mathcal{F}_{\D_n}=\bra{0_{AdS}}\sigma^{(bulk)}_{n}(y_1)\bar{\sigma}^{(bulk)}_{n}(y_2)\ket{0_{AdS}}.
\ee
Eventually, we have obtained
\be
-\left.\dfrac{\p}{\p n} \mathcal{F}_{\D_n}(u,v)\right|_{n\rightarrow1}=\dfrac{c}{6}\sigma_{min.}=E_W. \label{eq:cbew}
\ee
It is worth stressing that we have extracted the entanglement wedge cross section within the CFT framework. 

\subsection{Cross section of BTZ blackhole from conformal blocks}\label{subsec:btz}
Next, we derive the entanglement wedge cross section of BTZ blackhole \cite{Banados:1992wn} from the semi-classical Virasoro CBs. For simplicity, we will only consider the static case. The metric of static BTZ blackhole with mass $M$ is given by
\be
ds^2=\dfrac{\alpha^2}{\cos^2\rho}\left(\dfrac{d\rho^2}{\alpha^2}-dt^2+\sin^2\rho d\phi^2\right),\label{eq:sbtz}
\ee
where $\alpha^2$ is defined as
\be
\alpha^2\equiv-8G_NM<0.\label{eq:alpha}
\ee
It is well-known that the above metric can be obtained from the global AdS${}_3$ one \eqref{eq:gads3} with coordinate transformation $(t,\phi)\mapsto(\alpha t, \alpha \phi)$. 

\subsubsection{Similar phase as global AdS${}_3$}\label{subsubsec:btz1}
\begin{figure}[tbp]
\begin{center}
\resizebox{110mm}{!}{\includegraphics{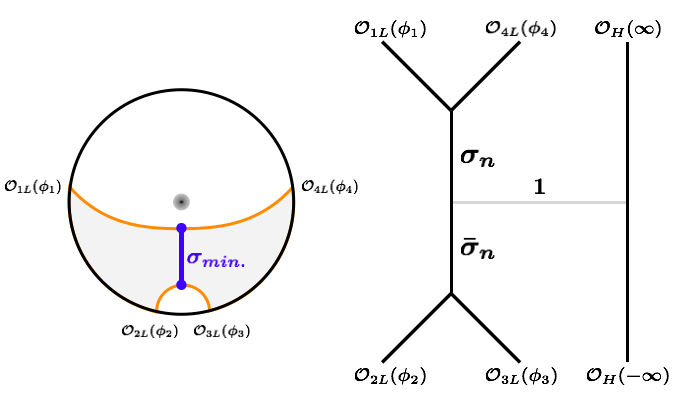}}
\caption{Left figure: a entanglement wedge (shaded region) for BTZ blackhole. In this situation, the wedge does not cover blackhole horizon (black point on center). Here we consider a timeslice $t=0$. Blue solid line represents the minimal cross section. Right figure: related OPE channel for the left diagram. Since we take the semi-classical heavy-light limit, the identity exchange becomes factorized. Hence, it reduces to the 4pt global CB with coordinates transformation $\phi^\prime=\alpha\phi$. }\label{fig:gwd2}
\end{center}
\end{figure}

Let us first consider the case when the entanglement wedge does not cover the blackhole (See left side of Figure \ref{fig:gwd2}).
In this case, we can follow the previous argument for global AdS${}_3$ and conclude that EoP can be obtained from the bulk-bulk propagator on the geodesics of BTZ blackhole. 
Hence, one can simply obtain EoP in this case from the coordinate transformation of \eqref{eq:cbew}. Moreover, this argument can be translated into the transformation of CB. Namely, we can use the fact that the heavy-light Virasoro CB can be obtained from the global CB in the previous subsection with coordinate transformation $(\phi^\prime,\tau^\prime)=(\alpha\phi,\alpha\tau)$ for the external light operators\cite{Fitzpatrick:2015zha}. 

In what follows, we will be more precise about our setup. We are now considering the semi-classical heavy-light Virasoro CBs (heavy-light CBs, for brevity), associated with the following 6pt correlator,
\be
\braket{0|\mathcal{O}_H(\infty)\mathcal{O}_H(-\infty)\mathcal{O}_{1L}(\phi_1)\mathcal{O}_{2L}(\phi_2)\mathcal{O}_{3L}(\phi_3)\mathcal{O}_{4L}(\phi_4)|0}.
\ee
Since we are considering the time slice of the boundary of \eqref{eq:sbtz}, we wrote each point of the cylinder as $\phi_i$. Here heavy operators $\mathcal{O}_H$ have scaling dimension $\D_H\sim c$ which can be identified with a pure state behaving like the BTZ blackhole\footnote{We can relate $\D_H$ to $\alpha$ such that $\alpha^2=1-\frac{12\D_H}{c}$. Here $c$ is the central charge identified with $c\equiv\frac{3}{2G_N}$ in the gravity side \cite{Brown:1986nw}.}. 
One could regard the state $\ket{\mathcal{O}_H}$ as a purification of the original thermal state. On the other hand, scaling dimension of light operator $\mathcal{O}_{iL}$ denoted by $\D_{iL}$ is small enough under the large-$c$ limit (again, we assume $\D_{iL}$s are the same). More precisely, we will assume $\D_{iL}/c\ll1$ so that $\mathcal{O}_{iL}$s can probe the BTZ geometry without any back reaction\cite{Asplund:2014coa}. Again, this condition will be  satisfied since the $\D_{n}$ becomes small under the limit $n \rightarrow 1$.  After the transformation and the semi-classical limit, all contribution of CB other than the global sectors become negligible at the leading order of the large-$c$ limit. 

This is generic argument for the heavy-light CBs, but let us focus on the OPE channel that $\mathcal{O}_H$s fuse into the identity operator (and its Virasoro descendants). Then, our heavy-light CB, say $\mathcal{F}^{(6)}_{\D_n}$, reduces to 4pt global CB with the coordinates transformation (times $\braket{\mathcal{O}_H(\infty)\mathcal{O}_H(-\infty)}$ that is normalized). We are interested in the case that the remaining two pairs of $\mathcal{O}_{iL}$s fuse into the twist operators. See right panel of Figure \ref{fig:gwd2}. After all, $\mathcal{F}^{(6)}_{\D_n}$ reduces to 4pt global CB on the new coordinates $\phi^\prime=\alpha\phi$. Evaluating \eqref{eq:gwdt} in the new coordinates with the saddle point approximation for $\D_n$, we have obtained 
\be
\mathcal{F}^{(6)}_{\D_n}(u,v)\sim e^{-\D_n\sigma_{min}(u,v)}, \label{eq:dombtz}
\ee
where $\sigma_{min}(u,v)$ matches the entanglement wedge cross section of BTZ geometry (blue solid line on the left panel of Figure \ref{fig:gwd2}, see also appendix \ref{app:A}). 
\subsubsection{A new phase: sections touching the horizon}\label{subsubsec:btz2}
\begin{figure}[tbp]
\begin{center}
\resizebox{100mm}{!}{\includegraphics{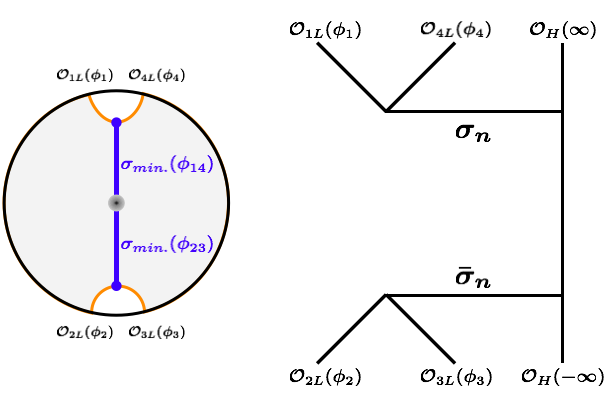}}
\caption{For sufficiently large subsystems, a entanglement wedge covers the blackhole horizon. In this case, the entanglement wedge cross section becomes blue solid lines $\sigma_{min.}(\phi_{14})+\sigma_{min.}(\phi_{23})$ in the left figure. Right figure shows the related OPE channel. Under the appropriate limit, the corresponding 6pt heavy-light CB becomes equivalent to product of two 4pt heavy-light CBs. Each of CBs produces the one of two cross section, $\sigma_{min.}(\phi_{14})$ or $\sigma_{min.}(\phi_{23})$.}\label{fig:gwd3}
\end{center}
\end{figure} 
There is another phase of the entanglement wedge that covers the horizon.  
In this case, rigorous GWD expression of 6pt CB has not been known yet. On the other hand, it is known that the heavy-light CBs can be obtained from the different method, called world-line approach\cite{Hijano:2015rla}(see also \cite{Alkalaev:2015wia,Alkalaev:2015lca,Banerjee:2016qca}). In particular, the 6pt heavy-light CB in Figure \ref{fig:gwd3} becomes product of 4pt ones\cite{Banerjee:2016qca},
\be
\mathcal{F}^{(6)}(\D_n, \D_H,c,\phi_i)=\mathcal{F}^{(4)}_{\D_n}(\D_H,c,\phi_{14})\mathcal{F}^{(4)}_{\D_n}(\D_H,c,\phi_{23}). \label{eq:644}
\ee
Here $\mathcal{F}^{(4)}_{\D_n}$s are 4pt heavy-light CBs\footnote{Our convention of the Virasoro conformal blocks is different from \cite{Banerjee:2016qca} and so forth. We are just using the terminology ``conformal blocks'' so that it depends only on the cross ratio.}. 
The above equality \eqref{eq:644} is verified only when the saddle point approximation can be applied,
\be
\mathcal{F}^{(4)}_{\D_n}(\D_L,\D_H,\alpha,\phi_{ij})\sim e^{-\D_n\sigma_{min}(\phi_{ij})}.
\ee
This situation is just what we want to consider. In the end, we get
\be
-\dfrac{\p}{\p n}\left. \left(\mathcal{F}^{(6)}(\D_n,\D_L,\D^\prime_L,\D_H,c,\phi_i)\right)\right|_{n\rightarrow1}=\dfrac{c}{6}\left(\sigma_{min}(\phi_{14})+\sigma_{min}(\phi_{23})\right), \label{eq:ewcth}
\ee
where
\be
\sigma_{min}(\phi_{ij})=\log\left|\dfrac{\cos\frac{\alpha\phi_{ij}}{4}}{\sin\frac{\alpha\phi_{ij}}{4}}\right|. 
\ee
The right hand side of \eqref{eq:ewcth} is nothing but the entanglement wedge cross section in left panel of Figure \ref{fig:gwd3}. 

Before closing this section, we briefly mention the case of the two sided eternal blackhole\cite{Maldacena:2001kr}. The bulk computation has been discussed in \cite{Nguyen:2017yqw}. If the two boundary regions, $A$ and $B$, are on the different side, then the entanglement wedge cross section is inhaled by the horizon and does not cover the entire blackhole. It is interesting to search its counterpart in the boundary, but we leave it for future work. 

\section{Summary and Discussion}\label{sec:discussion}
\ \ \ \ In this paper, we have proposed a formula \eqref{eq:eopcb} for EoP in two dimensional holographic CFT. We explained validity of \eqref{eq:eopcb} with the aid of the holographic code model in section \ref{sec:HaPPY}. Moreover, our formula reproduces the entanglement wedge cross section in a time slice of the AdS${}_3$ and one of the static BTZ blackhole. We observed this agreement in section \ref{sec:ew}. 

From the argument in section \ref{subsec:2pt}, we could not specify the external operators $\mathcal{O}_{iL}\,(i=1,2,3,4)$ in section \ref{sec:ew}, but at least the twist number must be conserved modulo $n$ in each OPE. Since the twist operator $\sigma_n (\bar{\sigma}_n)$ belongs to the twisted sector with twist number $\pm1$, one possibility might be that the $O_{iL}$ belongs to the sector with twist number $\pm\frac{n+1}{2}$, where $n$ is supposed to be odd integer before the analytic continuation of $n$. If so, the correlation function like \eqref{eq:4pt} could explain why the $\mathcal{O}(c)$ contribution of EoP vanishes under the transition from the entanglement wedge to the causal one. Namely, if we take another OPE channel with fixed external operators the above, internal twist operators cannot be produced  due to the twist number conservation. The relation between causal/entanglement wedge and OPE channels is reminiscent of the one for holographic mutual information. Identifying external operators in holographic CFT may provide a clue for EoP in more generic QFT. 

We focused on the two dimensional CFT and its bulk dual. One may be curious about its extension to the higher dimension. Since the twist operators on the higher dimension become non-local, generalization of our argument is not so straightforward. At least the $E_{P}=E_{W}$ conjecture can still work even in the higher dimension, so it will be fruitful to study the higher dimensional counterpart of our $\sigma^{(bulk)}_{n}$ in section \ref{sec:HaPPY}. 

Since our insight and optimization were based on the holographic code model, we are still assuming some holography. In particular, we do not say that our argument proves the $E_P=E_W$ even for holographic CFT. 
For further verification of \eqref{eq:eopcb}, we need to consider optimization within the framework of field theories. To this end, it would be very useful to utilize cMERA\cite{Haegeman:2011uy} or path integral approach like \cite{Caputa:2017urj}. However, at the very least, the right hand side of \eqref{eq:eopcb} defines a quantity of correlation measure in CFT that indeed agrees with the entanglement wedge cross section at the large-$c$ limit. Therefore, it would be very interesting to study further \eqref{eq:eopcb} with the $1/c$ corrections and its time dependence. 

It is also promising to see the connection of EoP to the kinematic space\cite{Czech:2015qta} since conformal blocks can be regarded as two point function of the OPE blocks\cite{Czech:2016xec, deBoer:2016pqk}. We leave these questions as future work. 

\bigskip
\goodbreak
\centerline{\bf Acknowledgments}
\noindent
We are grateful to Tokiro Numasawa, Tadashi Takayanagi, and Satoshi Yamaguchi for valuable comments and discussion. KT would like to thank the workshop ``Holography, Quantum Entanglement and Higher Spin Gravity II'' where a part of this work was presented.

\appendix
\section{Explicit form of propagator and minimal geodesic length}\label{app:A}
\ \ \ \ In this appendix, we note some explicit form of the quantities displayed in section \ref{sec:ew}.
The bulk-bulk propagator $G^{\D_p}_{bb}(y,y^\prime)$ (in AdS${}_3$) is
\be
G^{\D_p}_{bb}(y,y^\prime)=\dfrac{e^{-\D_p\sigma(y,y^\prime)}}{1-e^{-2\sigma(y,y^\prime)}}. 
\ee
Here $\sigma(y,y^\prime)$ is the geodesic distance between $y$ and $y^\prime$,
\be
\sigma(y,y^\prime)=\log\left(\dfrac{1+\sqrt{1-\xi^2}}{\xi}\right), \;\xi=\dfrac{\cos\rho\cos\rho^\prime}{\cos(t-t^\prime)-\sin\rho\sin\rho^\prime\cos(\phi-\phi^\prime)},
\ee
where we are taking the global coordinates \eqref{eq:gads3}. 
Geodesics $\g_{ij}$ anchored on the boundary points $(\phi_{i},t_i=0)$ and $(\phi_j,t_j=0)$ are given by
\begin{subequations}
\begin{align}
\cos\rho(\la)&=\dfrac{\big|\sin\frac{\phi_{ij}}{2}\big|}{\cosh\la},\\
e^{2i\phi(\la)}&=\dfrac{\cosh\big(\la-\frac{i\phi_{ij}}{2}\big)}{\cosh\big(\la+\frac{i\phi_{ij}}{2}\big)}e^{i(\phi_i+\phi_j)}, 
\end{align}
\end{subequations}
where $\la$ is a proper distance for the geodesic and $\phi_{ij}=\phi_i-\phi_j$.

The minimum length between two geodesics $\sigma_{min}$ in \eqref{eq:dom} is given by the cross ratio $u,v$,
\be
\sigma_{min}(u,v)=\log\left(\dfrac{1+\sqrt{u}+\sqrt{(1+\sqrt{u})^2-v}}{\sqrt{v}}\right),
\ee
where
\be
u=\dfrac{P_{12}P_{34}}{P_{13}P_{24}},\;\;\; v=\dfrac{P_{23}P_{14}}{P_{13}P_{24}}.
\ee
Here $P_{ij}$ (on a time slice $t=0$) is 
\be
P_{ij}=4\sin^2\left(\frac{\phi_{ij}}{2}\right),
\ee
for global AdS${}_3$. One can also obtain $\sigma_{min}$ for static BTZ blackholes in section \ref{subsubsec:btz1} by replacing $P_{ij}$ with $P^{(\alpha)}_{ij}$,
\be
P^{(\alpha)}_{ij}=4\sin^2\left(\frac{\alpha}{2}\phi_{ij}\right), 
\ee
where $\alpha$ is pure imaginary number defined in \eqref{eq:alpha}.

\end{document}